\documentclass[12pt,a4paper]{article}
\usepackage{graphicx}

 \newcommand{\bc}{\begin{center}}
 \newcommand{\ec}{\end{center}}
                   \newcommand{\bfr}{\begin{flushright}}
                   \newcommand{\efr}{\end{flushright}}
   \newcommand{\ii}{\item}
     \newcommand{\be}{\begin{enumerate}}
     \newcommand{\ee}{\end{enumerate}}
        \newcommand{\bi}{\begin{itemize}}
        \newcommand{\ei}{\end{itemize}}
            \newcommand{\bd}{\begin{description}}
            \newcommand{\ed}{\end{description}}
                \newcommand{\beq}{\begin{equation}}
                \newcommand{\eeq}{\end{equation}}
                  \newcommand{\bea}{\begin{eqnarray}}
                  \newcommand{\eea}{\end{eqnarray}}

      \newcommand{\bfi}{\begin{figure}}
      \newcommand{\efi}{\end{figure}}
\newcommand{\bay}{\begin{array}{l}}
\newcommand{\eay}{\end{array}}
            \newcommand{\dd}{\mbox{d}}

    \newcommand{\Del}{\Delta}
    
    \newcommand{\la}{\lambda}

    \newcommand{\eps}{\epsilon}
    \newcommand{\tht}{\theta}  

    \newcommand{\om}{\omega}

    \newcommand{\vf}{\varphi}

\begin{document}   \vskip 1.5in
\thispagestyle{empty}
\hspace*{1mm}  \vspace*{-0mm}
\noindent {\footnotesize {{\em
			\hfill      Posted on ArXiv, submitted to Materials and Structures} }}
\vskip 1in
\begin{center}
		{\Large {\bf 
		Statistical Extraction of Useful Concrete Creep Data from 
		Imperfect Customary Tests
}}\\[20mm]
	
	{\large {\sc Mohammad Rasoolinejad, Saeed Rahimi-Agham, and Zden\v ek P. Ba\v zant }}
	\\[1in]
	
	{\sf SEGIM Report No. 18-07/788m}\\[1.5in]
	
	Center for Structural Engineering of Geological and Infrastructure Materials (SEGIM) 
	\\ Department of Civil and Environmental Engineering
	\\ Northwestern University
	\\ Evanston, Illinois 60208, USA
	\\[1in]  {\bf July, 2018} 
\end{center}

\clearpage   \pagestyle{plain} \setcounter{page}{1}

\title{Extended Microprestress-Solidification Theory (XMPS) for Long-Term Creep and Diffusion Size Effect in Concrete at Variable Environment}
\author{{Saeed Rahimi-Aghdam} \\
	{\aff{Graduate Research Assistant, Northwestern University, Evanston, IL.}} \\
	\\
	{\authornext{Zden\v ek P. Ba\v zant}}\\
	{\aff{McCormick Institute Professor and W.P. Murphy Professor of Civil
			and Mechanical Engineering }}\\
	{\aff{and Materials Science, Northwestern University 2145 Sheridan Road, CEE/A135, Evanston, Illinois}} \\
	{\aff{60208; z-bazant@northwestern.edu; corresponding author.}} \\ 
	\\
	{\authornext{Gianluca Cusatis}}\\
	{\aff{Associate Professor of Civil and Environmental Engineering, Evanston, IL.}}}
\date{}
	\begin{center}
		{\Large {\bf 
				Statistical Extraction of Useful Concrete Creep Data from 
				Imperfect Customary Tests
		}} 
		\\[7mm]  {\large Mohammad Rasoolinejad, Saeed Rahimi-Aghdam and Zden\v ek P. Ba\v zant}

	\end{center}  
	
\noindent {\bf Abstract:}\, {The reporting and evaluation of creep tests of concrete is complicated by the fact that creep is significant even for the shortest observable load durations. Compared to the strain after 0.1 s load duration, the strain at 2 hour duration is typically 53\% greater. Most experimenters have for decades been unaware of this fact. Consequently, the reported creep curves require correction by a time shift, which ranges from 0 to 2 hours. This further implies a vertical shift of entire creep curve, important for all times up to structure lifetime.
	To filter out the errors, it is argued that, within an initial period during which the advance of hydration is negligible, which is normally about 1 day, the initial basic creep must follow a power law of the time. Creep test data from the literature are used to prove it. Corrections by time and deformation shifts are determined by minimization of the sum of squared deviations of the power law from the creep test data. For a fixed exponent $n$ and time shift $s$, the optimization is reduced to linear regressions of two kinds, depending on whether the data are given in terms of either the compliance function or the creep coefficient. For both, the linear regression parameters depend nonlinearly on the chosen values of $n$ and $s$. To avoid nonlinear optimization, which need not converge to the correct result, a set of many discrete values of $n$ and $s$ within their realistic ranges is selected and the $(n,s)$ combination minimizing the objective function is obtained by a search. Enforcing a power law form of the initial creep curve is found to lead to better data fits. The optimum exponent $n$ for the entire database is around 0.3, applicable to the time period cca (10 s, 1 day). After that, the exponent transits to about 0.1, and prior to that it is about 0.08. After filtering out the errors, the corrected database will allow better calibration of the general creep prediction model such as B3 or B4.       }
	


\section*{Introduction}
\label{intro}
In most concrete creep testing, it has been customary to report either the creep coefficient $\vf$, representing the creep-to-elastic strain ratio, or the compliance function determined from the measured creep part of strain. Unfortunately, though, information is missing on 1) the rate and evolution of the load as the load is raised to the sustained load level, 2) the moment of zero time for the reported creep data, 3) the magnitude of initial total deformation, and 4) the time corresponding to this deformation. As it turns out, the initial uncertainties can lead to large errors for the entire load duration. Such errors, unfortunately, contaminate virtually all the existing data, including the latest worldwide Northwestern University (NU) database containing over 4000 creep and shrinkage tests \cite{bazant2008comprehensive}, which was used to calibrate creep and shrinkage prediction model B4 \cite{hubler2015comprehensive} (the database is downloadable from http://www.civil.northwestern.edu/people/bazant or http://www.baunat.boku.ac.at/creep.html).

Proposed here is a statistical optimization method to filter out these errors and thus decontaminate the database. The method is based on the hypothesis that the initial creep curve ought to be a power law. This hypothesis is here justified physically.

\subsection*{Uncertainties in Initial Deformation}
\label{sub1:intro}
\begin{figure*}
	\centering
	\includegraphics[width=0.5\textwidth]{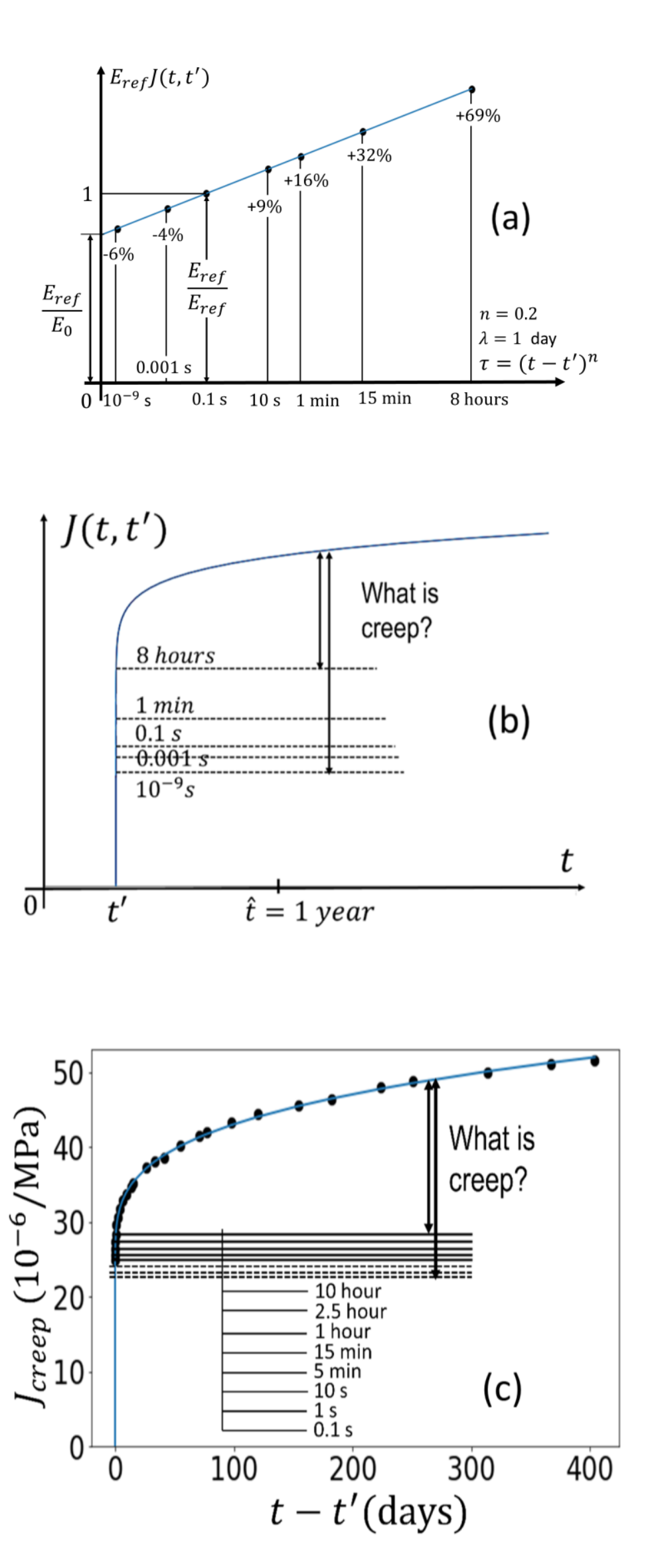}
	\caption{Compliance curve (for $n$=0.2) plotted in (a) transformed time scale, and (b) linear time scale; (b,c) uncertainties in creep values shows (b) schematically and in (c) in a plot of Kommendant et al.'s \cite{kommendant1976study} data (these data were found to be perfect, needing non correction)}
	\label{fig:linear & transformed}
\end{figure*}

The widespread habit of plotting the creep curves in a linear, rather than logarithmic, time scale (Fig.~\ref{fig:linear & transformed}) has caused significant errors. The linear scale plot creates an illusion that concrete creep begins only after about several hours of loading. In reality, it begins in much less than a microsecond after instantaneous load application, which cannot be discerned in a linear scale plot. The hydraulic creep test frames (as well as standard testing machines) can measure the deformation for 0.1 s duration. For durations less than a microsecond the evidence comes from measurements of the viscoelastic complex modulus in high frequency vibrations, and of the velocity of elastic shock waves or ultrasound.

The compliance function $J(t,t')$ is defined as the total strain at concrete age $t$ produced by a unit sustained stress applied at age $t'$, and thus includes the initial instantaneous elastic strain, $1/E_1$. In theory the load application should be sudden, in the form of Heaviside step function, but this is never followed in practice. Anyway, to ramp us the load without inertia effects in $<0.1$ is next to impossible. The creep actually begins already within less than $<10^{-9}$ s after sudden load application and, typically, the strains obtained within 0.1 s and 1 hour after a sudden load application differ by 30\% to 40\%. The practical testing has deviated from the ideal step function in four different ways:
 \be
 \ii
The load is always ramped up gradually---at a uniform rate, or a variable rate, or in steps (Fig. \ref{fig:Shift}a). Concrete creep occurs already during the rise of the load. There exists a certain equivalent time for which a sudden step function loading would give about the same creep once the sustained load level is reached, but this equivalent time is not known, and is likely to be shifted compared to the reported times. This time shift, $s$, is, of course, negligible compared to the load durations of many years, but the problem is that it causes an error in evaluating the initial deformation, which, in turn, leads to a vertical shift $\Del$ of the entire compliance curve.
 \ii
Another error in the reported compliance function has often been introduced when the initial, supposedly elastic, deformation was measured by a separate test, e.g., the standardized test of elastic modulus $E$, and not in the creep test itself. Such initial errors cause the entire compliance-time curve to shift, even by up to 50\% of the initial instantaneous strain. This causes a similar error in the reported creep coefficient $\vf$. 
 \ii
Sometimes the $E$-value at the time $t_1$ of loading was not measured at all but was inferred from its relation to the compression strength, $f_c$, of the concrete, using the empirical ACI equation, given as Eq. A-39 in ACI-R209 \cite{videla2008guide}). This estimate, though, involves significant error. Further error in the $E$-value at age $t_0$ at load application arises when the standard value at age 28 days is used. In that case, another empirical equation, Eq. A-34 in ACI-R209, describing the concrete age effect on $E$, is used. This equation is also approximate and introduces an additional error.
 \ii
Many reports presented the creep test result only in the form of a plot in a linear time scale. Then, if the time scale ends at 1 hour, 1 day, 1 month, or 1 year, it is impossible to distinguish visually loading durations less than 10 s, 10 min, 2 hours or 1 day, respectively. This introduces a big uncertainty regarding the actual time of loading as well as the initial deformation.
 \ee

Do these errors justify discarding most of the existing big data on creep? Of course not. Over the last eight decades, enormous funds and human effort were invested into worldwide creep testing. The worldwide creep test data, which are now collected in the NU database, comprise over 4,000 long-time tests \cite{bazant2008comprehensive}. So, what is needed is to filter out the initial errors contaminating the database. The filtered database should allow a significant improvement in the predictive capability of a multi-decade creep prediction model such as B4 \cite{bazant2015rilem} (which became a RILEM recommendation).

The short-time evolution of $J(t,t')$ is, of course, unimportant for predicting multi-year and multi-decade creep effects in structures. But the problem is that the error in the initial value shifts vertically the entire compliance curve. The shift can be up to 50\% of the initial compliance.

It is now clear \cite{bavzant1995justification,wendner2015statistical,hubler2015statistical,wendner2015optimization} that the initial creep curve follows approximately the power law $(t - t')^n$, where $t-t'$ = load duration. So, after subtraction of the correct instantaneous strain, the logarithm of creep strain plotted vs. log$(t-t')$ must give a straight line of slope $n$. The slope can vary greatly but is typically between 0.05 and 0.40. It is instructive to look at numbers. Assume slope 0.2 and take the classical wisdom that the creep coefficient from 1 day to 1 year is around 2 (which gives $\la\approx$ 1 day). 
Then, for load durations of 0.1 s, 1 s, 1 min. and 1 hour, the total strains, i.e., $J(t,t')$, typically increases by about 6.5\%, 10\%, 23\% and 53\%, respectively, compared to the truly instantaneous strain, which occurs at load duration 0, virtually at $<10^{-9}$ s.    

Compared to the total strain at 0.1 s (which the duration load rises when a valve is opened in a hydraulic system), the strains at 1 min. and 1 hour are, respectively, about 16\% and 44\% larger (Fig.~\ref{fig:linear & transformed}a). Compared to the total strain at 1 min., the total strain at 1 hour is approximately 24\% larger. The 7-day creep coefficient $\vf$ is about 1.24 if related to the total strain at 1 s, and 0.81 if related to the total strain at 10 min. Relating the measured creep strain to the elastic deformation calculated on the basis of the standardized test of elastic modulus E, or estimated on the basis of compression strength $f_c$, gives again rather different values.

These discrepancies are quite significant. They have plagued creep analysis of structures for decades. Although the creep coefficient is convenient for approximate calculation of the structural creep effects in the traditional way, without a computer, it should always be evaluated from the directly measured compliance function $J(t,t')$. The short-time values of $J(t,t')$, for less than a day, are unimportant for long-term creep effects in structures, but they can lead to a big vertical shift of the compliance curve, which matters for all times. Consequently, the testing standards for creep should explicitly state that reporting the test results in terms of the creep coefficient, as well as determining the initial deformation from tests other than the creep test itself, is an erroneous and misleading practice.

The Northwestern University (NU) world-wide database \cite{bazant2008comprehensive,hubler2015comprehensive} contains data from 1433 creep tests, most of which suffer from this kind of errors. In previous studies, the errors in the initial strain definition were simply ignored. However, regardless of these errors, the NU database contains a wealth of information, which has been obtained at enormous cumulative cost, and cannot be discarded.

So the problem at hand is how to filter out the error from the reported creep data, at least approximately, and generate in a systematic way a filtered database of concrete creep. This filtering is not trivial, which is probably why it has not yet been done. But it is not difficult either, if the recent understanding of early creep is exploited and modern computer optimization techniques are utilized. Therefore, formulating a method to filter the creep database is adopted as the objective of this paper.

\subsection*{Basic Hypothesis: Physical Justification of Power-Law }
\label{sub2:intro}

Let us denote $\tht = t-t'$ = duration of sustained load. We restrict consideration to sufficiently short load durations $\Del \tht_{ini}$ after initial loading, during which the material changes due to hydration (or aging) or other phenomena are negligible. In practice, $\Del \tht_{ini}$ can be up to about 5\% of the age at loading, and in the case of simultaneous drying up to 5\% of the prior exposure to constant drying environment. During $\Del \tht_{ini}$, the creep process (caused by breakage of interatomic bonds at highly stressed creep sites and their restorations at adjacent sites) must be self-similar. Mathematically this requires the creep strains $f(\tht)$ and $f(\tht^*)$ at any two times $\tht$ and $\tht^*$ under constant load must satisfy the relation
 \beq  \label{s1}
  \frac{f(\tht)}{f(\tht^*)} = f \left(\frac \tht {\tht^*} \right)
 \eeq
which means that the response does not depend on the choice of reference time $t^*$. This is a functional equation, which happens to be easily solved. It can be checked by substitution that the power law with a non-zero exponent $n$:
 \beq  \label{s2}
  f(\tht) = \tht^n~~~~~(n \ne 0)
 \eeq
is a solution. Further it can be shown that the power law is the only solution \cite[Sec.9.4, p.417]{bavzant2018creep}.

What is the proper choice of the short-time range, in which self-similarity applies and the creep follows the power law? Examination of test data indicates that it may be chosen as
 \beq  \label{tt}
  t - t'\, \le 0.05\, t'~~~~\mbox{for $t' \ge 1$ week}
 \eeq
During this period, the advance of hydration as well as autogenous shrinkage is negligible and cannot significantly affect the creep.
This is confirmed by the evidence of power law creep in several datasets with many readings during the aforementioned interval (especially the tests of Kommendant et al.\cite{kommendant1976study}). In some data sets, there is in this range only one data point, or none, and for those, of course, no filtering of errors is possible. The restriction $t' \ge 1$ week is added to avoid $t - t'$ being too short.  

It may be noted that the power law character of initial creep appears applicable for a short time even when a specimen is exposed to drying simultaneously with loading. But then exponent $n$ may be considerably different.

\section*{Statistical Filtering of Imperfect Data via Optimization}
\label{section2}
\subsection*{Imperfect creep coefficient}
\label{section2:sub1}

Based on the aforementioned, physically justified, hypothesis, we assume that, for some initial period, the total strain in creep tests grows as $\eps  \propto \tau$ + constant where
 \beq \label{e1}
  \tau(n, s) = \tht^n, ~~~~  \tht = t - t' + s
 \eeq
Here $s$ is a certain time shift, not necessarily positive, and exponent $n$ is an empirical constant. While, in models B3 and B4, $n = 0.10$ was approximately the overall optimum for the whole database, in individual data, the optimal $n$ can vary widely. For many data sets it is not clear whether the beginning of time was taken as the moment at which the full sustained load value was reached, or as another moment during the load application process. 
Optimally, one should integrate the creep along the loading ramp and find for which time a sudden step-wise load application would give the same initial deformation under full load. An estimated creep law with no aging would suffice for this purpose, but the problem is that the ramp duration and history are unknown for all the data.

\begin{figure*}
	\centering
	\includegraphics[width=0.8\textwidth]{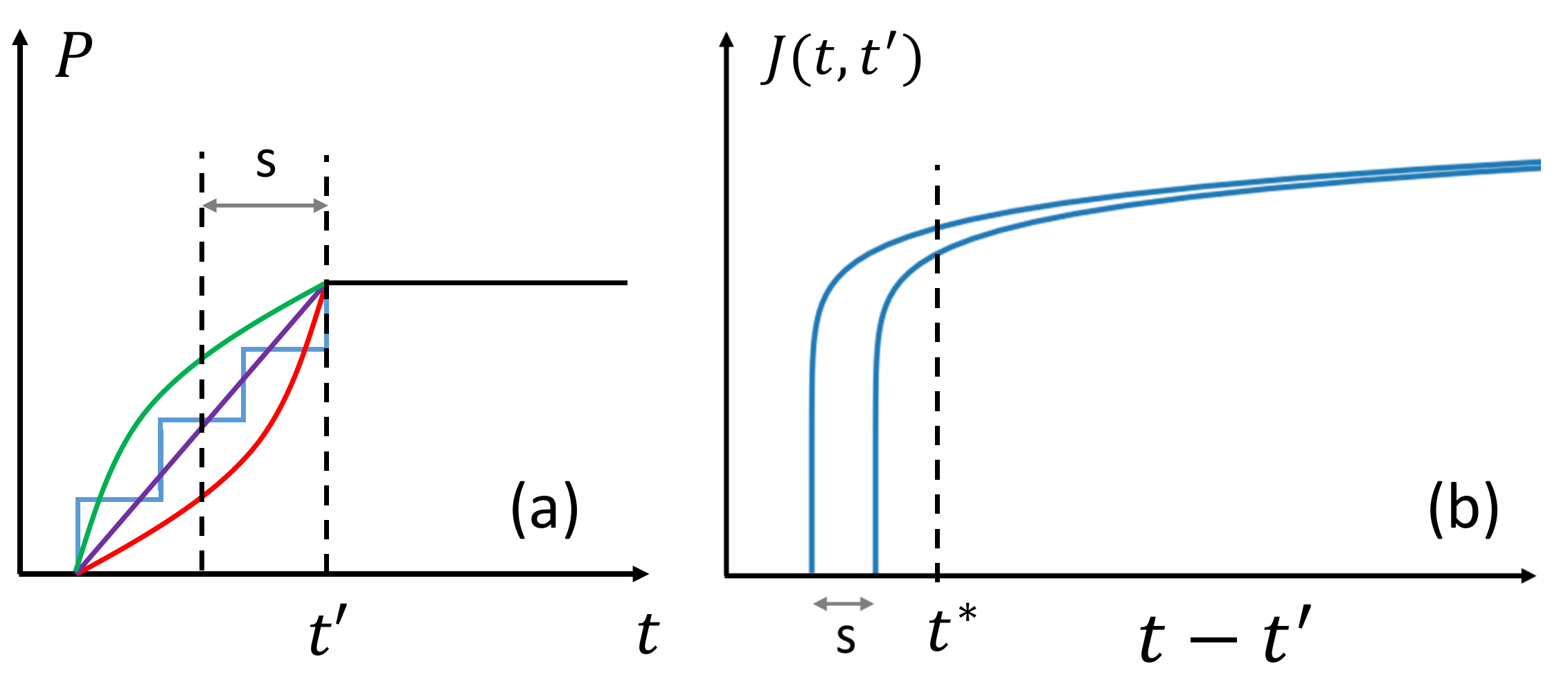}
	\caption{(a) Possible histories of initial load application (b) and their effect on creep curve}
	\label{fig:Shift}
\end{figure*}

Physically, shift $s$ yields the loading time for which a sudden load application with a step function load history would give the same creep curve as the gradual load application in the actual test; see the sketch in Fig. \ref{fig:Shift}. Actually, this condition can be satisfied accurately only after a certain lapse of time, and this is a weakness of the assumption that the initial creep curve is a power law. It could be exactly a power law only for step function loading. In this regard, if the process of applying the load takes time $\Del t$, one should better discard data points, if any, during a time interval equal to $\Del t$ after the instant at which the load reached the full value of sustained load (Fig. \ref{fig:Shift}). The problem, of course, is that time $\Del t$ is generally not reported. But this is a simplification that we have to accept.

It may be noted that recently some researchers \cite{vandamme2013creep} interpreted their micro-indentation creep tests during the initial few seconds as approximately logarithmic, which would correspond to $n \to 0$; but their data were later found to be fitted slightly better by a power law with $n = 0.08$ (see Fig. 5 in \cite{wendner2015statistical}), although visually the difference is small.

\begin{figure*}
	\centering
	\includegraphics[width=0.8\textwidth]{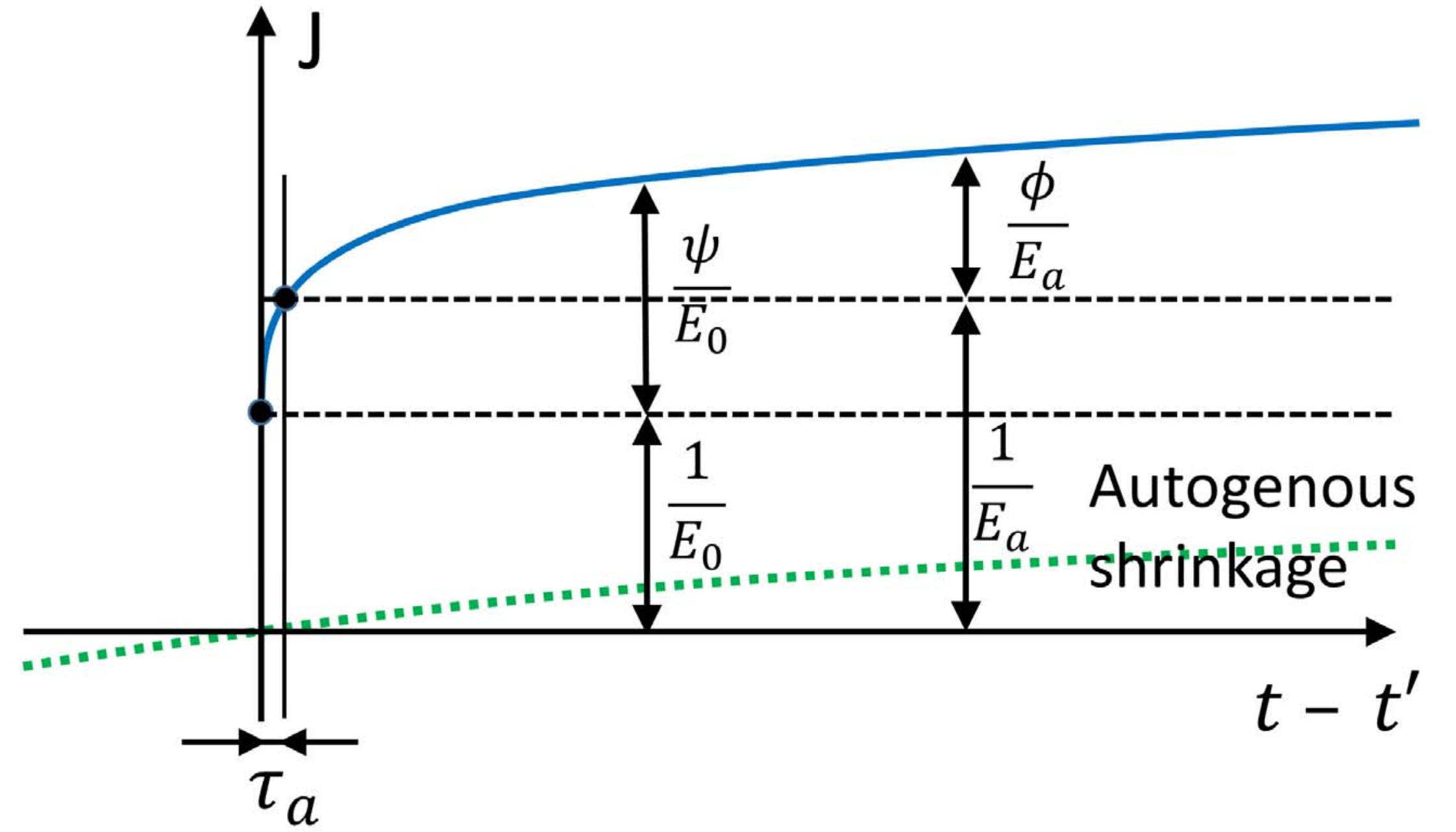}
	\caption{True and apparent creep coefficient in compliance function}
	\label{fig:schematic}
\end{figure*}

The compliance function of concrete may be written as
 \bea \label{e2}
  J(t,t') &=& \frac{ 1 + \psi(t,t')}{ E_0 }
 \\ \label{e2a}
  \psi(t,t') &=& \frac{\tau(n, s)}{\la^n}
 \eea
where $E_0$ is the truly instantaneous elastic modulus
(while $1 /E_0$ is the truly instantaneous elastic deformation) (Fig. \ref{fig:schematic}); $\psi(t,t')$ is the true creep coefficient; $\la$ is a factor (of time dimension) that is to be determined by creep data fitting; $\la$ makes $\psi(t,t')$ dimensionless; and $\la^{-n}$ represents a scaling factor of the creep coefficient referred to a zero load duration, which cannot be determined if the value of $E_0$ has not yet been identified.

In terms of the conventional (or apparent) creep coefficient $\vf$, the compliance function is
 \beq \label{e3}
   J(t,t') = \frac{ 1 + \vf(t,t')}{ E_a }
 \eeq
where $E_a$ it the reported apparent elastic modulus, defined in one way or another by the experimenter (see Fig. \ref{fig:schematic}) in disregard of the short-time uncertainties discussed here.

\begin{figure*}
	\centering
	\includegraphics[width=0.99\textwidth]{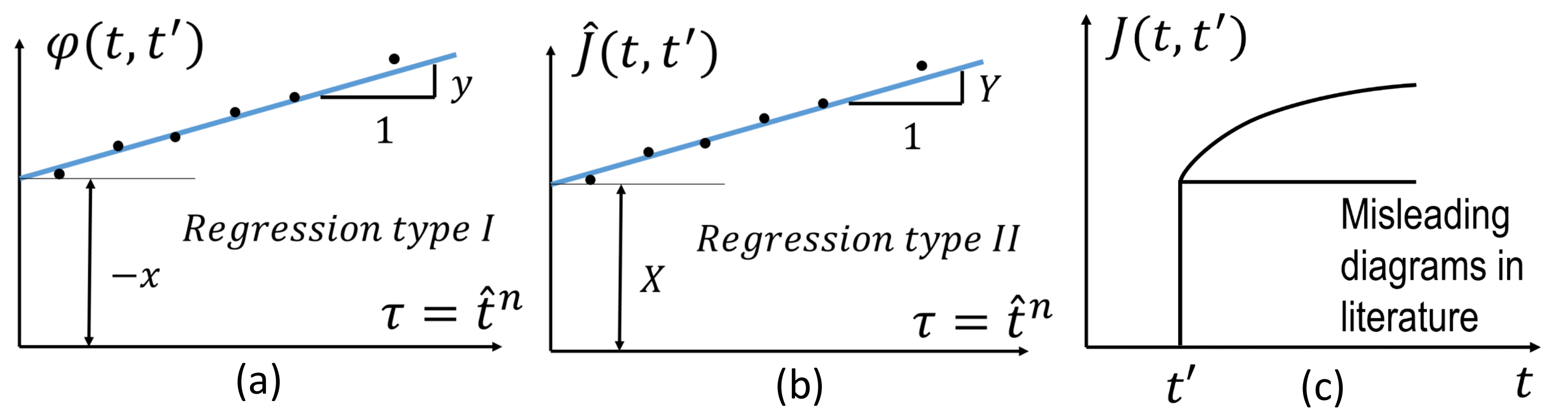}
	\caption{Parameters of creep curve regressions for (a) creep coefficient and (b) creep compliance; (c) misleading schematic of concrete creep curves found in literature}
	\label{fig:fitting}
\end{figure*}

Equating the right-hand sides of Eqs. (\ref{e2}) and (\ref{e3}), we get
 \beq  \label{e4}
  \vf(t,t') = \frac{E_a}{E_0}\, [1 + \la^{-n} \tau(n,s)] - 1
 \eeq
Let us now set
 \bea \label{e5}
  x &=& 1 - \frac{E_a}{E_0} 
 \\ \label{e5a}
  y &=& \frac{E_a}{\la^n E_0}
 \eea
where $E_a$ is the value retrieved from the experimenter's report. With this notation, Eq. (\ref{e4}) becomes:
 \beq \label{e7}
  - x + \tau(n,s) y = \vf(t,t')
 \eeq
If $n$ and $s$ are specified, this a linear regression equation, in which $y$ is the regression slope and $-x$ is the regression line intercept (Fig. \ref{fig:fitting}a). The $\vf(t,t')$ values expressed as a function of the transformed measurement times $\tau_i$ ($i = 1, 2, 3,...N$) represent the data.

Once the optimum $x$ and $y$ values are obtained, the inverses of Eqs. (\ref{e5}) and (\ref{e5a}) give
 \bea \label{e6}
  E_0 &=& \frac{E_a}{1-x}   
 \\ \label{e6a}
  \la^n &=& \frac{E_a}{E_0 y} = \frac{1-x} y
 \eea
where the regression parameters $x$ and $y$ depend of the chosen $n$ and $s$. According to Eq. (\ref{e2}), $1/E_a = (1 + \la^{-n}\tau_a)/E_0$ or $\tau_a = \la^n [ (E_0 /E_a) - 1 ]$. This yields
 \beq \label{e8}
  \tau_a(x,y,s) = \frac x y
 \eeq
If $n$ and $s$ are known, this is the optimum estimate of the unreported loading time (or duration) that corresponds to the reported $E_a$ value.


The regression based on Eq. (\ref{e7}) minimizes the sum
 \beq \label{s11}
  F(x,y,n,s) = \sum_{i=1}^{N_i} [\tau_i(n,s) y - x - {\vf_a}_i]^2
 \eeq
where ${{\vf_i}_a}_i$ are the apparent creep coefficient values corresponding times $(t-t')_i$ as reported by the experimenter. Because $\tau_i(n)$ depends on the unknown exponent $n$ and time shift $s$, the minimization of $F(x,y,n,s)$ is a nonlinear optimization problem. This problem is complicated by the need to impose restrictions on the realistic ranges of $n, s$ and $\tau$.

Because of a large possible mismatch in the reported $E_a $ and $\vf$ values, as well as scarcity of sufficiently short times among the creep data, the linear regression can unfortunately produce values of $n$ and $\tau$ that fall outside reasonable bounds. If, for instance, the $n$ value optimized for some data set could yield a $\tau_a$ value lying outside the interval (1 s, 4 hours) for a spring-loaded test frame, or outside (0.001 s, 5 min.) for a hydraulic test setup, it would be an optimization error attributable to the random scatter of data or to scarcity of short-time readings.

To prevent such optimization outcomes, limits on both $n$ and $\tau_a$ need to be imposed. In particular, $n$ values exceeding 0.40 are unreasonable, and so are the $\tau_a$ values outside the range
 \beq  \label{tau range}
  {\tau_a}_{\,min} \le \tau \le {\tau_a}_{\,max}
 \eeq

So, a restricted optimization is necessary. A robust and effective way to implement the restrictions is a discretized optimization:
 \be
 \ii
Choose a set of discrete, closely spaced, $n$ and $s$-values such as 
 \bea   
 \label{nn}
  && n = \, 0.01, 0.02, 0.03,..., n_{max},~~~~n_{max} = 0.40
 \\ 
  \label{ss}
  && s = 
  \, 0, \pm 0.1, \pm 0.3, \pm 1, \pm 3, \pm 10, \pm 30, \pm 100, \pm 300, \pm 1000,     
  \pm 3000, \pm 10000~s
  \eea   
Even though the database indicates no exponent $n$ to be less than 0.05, the optimization covers $n$ values down to 0.01. The reason is that recent nano-indentation tests \cite{vandamme2013creep} of about 10 s duration indicated exponent $n$ close to 0, and some researchers think that this exponent applies for multi-decade creep, too. The fact that the database optimization yielded the average $n$ to be about 0.3 shows that the ten-second nano-indentation test cannot be extended to hourly durations.

Also, choose a set of discrete, closely spaced, $\tau_a$-values within the interval
 \beq \label{tau}
  {\tau_a}_{\,min} \le \tau_q \le {\tau_a}_{\,max}     
 \eeq
where one may typically assume
 \beq \label{n'}
  {\tau_a}_{\,min} = 1~\mbox{s},~~~{\tau_a}_{\,max} = 4~\mbox{hours}
 \eeq
 \ii
For each $n, s$ and $\tau_a$, a linear regression subroutine based on Eq. (\ref{e7}) furnishes $x$.
 \ii  
Calculating for each case $E_a = (1-x)E_0$, one may now exclude the cases for which $E_0$ lies outside some chosen reasonable interval $({E_0}_{\,min}, {E_0}_{\,max})$, although this step is needed only if one desires more stringent limits on $E_0$ than those imposed indirectly by the chosen limits on $\tau_0$ according to equations $x = y \tau_a$ and $E_0 = E_a /(1-x)$.
 \ii
For each $n$ and $s$ and the corresponding $\tau_a$, evaluate the sum of squares $F(x,y,n)$.
 \ii
Finally, search for the $(n,s)$ combination that yields the smallest $F(x,y,n,s)$.
 \ee

\subsection*{Imperfect Compliance Function}
\label{section2:sub2}

Properly, the creep should be characterized in terms of the compliance function $J(t,t')$ which includes the initial instantaneous (or elastic) deformation.
However, the time shift, $s$, may have led to some vertically shifted compliance function $\hat J(t,t')$. Besides, what was sometimes reported may have been a compliance function constructed either from $\vf(t,t')$ or from a measured strain increase after the initial strain reading, using an imprecisely defined elastic modulus $\tilde E$. E.g., the value of $\tilde E$ may have been obtained by a separate short-time test. It may have been determined by the standardized test of elastic modulus, or estimated from the strength of concrete. Such imperfections may be revealed by a significant deviation of the initial period of the curve of $J(t,t')$ versus $t$ from the power law, $(t-t')^n$ + constant.

From Eqs. (\ref{e2}) and (\ref{e2a}), defining creep coefficient $\psi(t,t')$ in terms of instantaneous modulus $E_0$, we have
 \beq \label{e11}
  \psi(t,t') = \tau(n,s) \la^{-n}~~~\mbox{and}~~~
  \psi(t,t') =  E_0 \hat J(t,t') - 1
 \eeq
This may be rearranged as
 \beq \label{e12}
  X + \tau(n,s) Y = \hat J(\tau)
 \eeq
For fixed $n$ and $s$, this is a linear regression relation (Fig. \ref{fig:fitting}b) in which the intercept and slope of the regression line are
 \beq \label{e13}
  X = \frac 1 {E_0}, ~~~~ Y = \frac 1 {\la^n E_0}
 \eeq
where $X$ and $Y$ depend on $n$ and $s$. Once the values of $X$ and $Y$ are solved, one can invert these relations to evaluate:
  \beq \label{e14}
   E_0 = \frac 1 X, ~~~~ \la^n = \frac X Y
  \eeq
For any fixed $n$, the linear regression minimizes the sum
 \beq  \label{e15}
  F(X,Y,n,s) = \sum_{i=1}^{N_i} [\tau_i(n,s) Y + X - \hat J_i ]^2
 \eeq

Similar as before, the scatter or scarcity of the short-time data may cause the $E_0$ value delivered by linear regression to be outside a reasonable range, defined as
 \beq \label{e12a}
  {E_0}_{\,min} \le E_0 \le {E_0}_{\,max}
 \eeq
where it might be suitable to assume
 \beq \label{e12c}
  {E_0}_{\,min} = 1.1 E_a,~~~ {E_0}_{\,max} = 5 E_a.
 \eeq
The reason for setting the lower bound on $E_0/E$ as 1.1 is that the asymptotic initial strain is, and must always be, appreciably less than the measured Young's modulus. The reason for the upper bound of 5 is that the sub-millisecond creep needed to explain a larger value would be unreasonably large.           

\begin{figure*}
	\centering
	\includegraphics[width=0.5\textwidth]{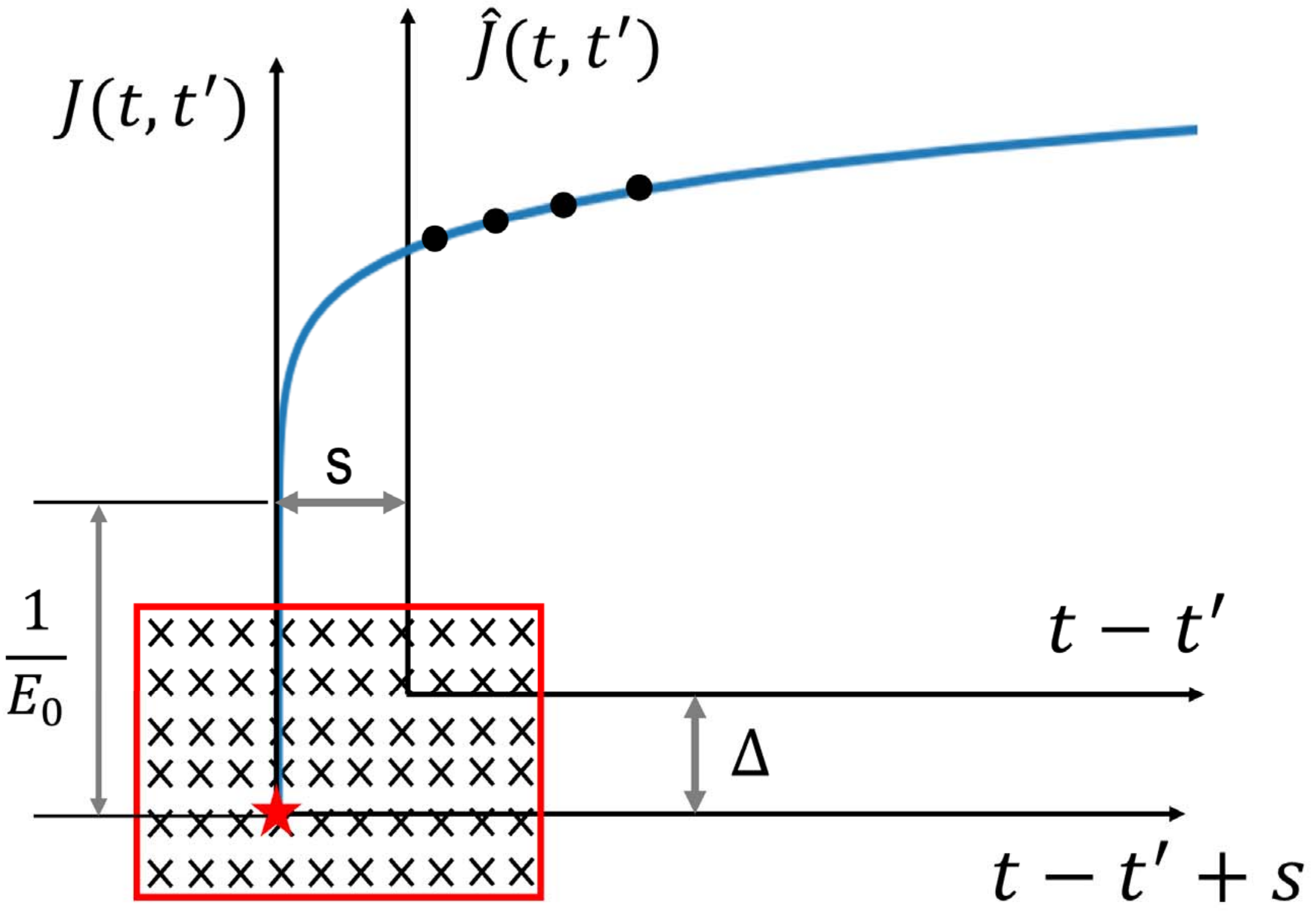}
	\caption{Schematic of constrained optimization showing various possible locations of coordinate origin for various horizontal shifts $s$ and vertical shifts $\Del$ of the compliance curve}
	\label{fig:optimization}
\end{figure*}

To satisfy these limits, a discrete optimization procedure is pursued as follows (Fig. \ref{fig:optimization}):
 \be
 \ii
Choose a set of discrete closely spaced $n$-values and $s$ values such as
 \begin{equation}
\label{nn2}
n = \, 0.01, 0.02, 0.03,..., n_{max},~~~~n_{max} = 0.40
\end{equation}

\begin{equation}
\label{s'}
s = \, 0, \pm 0.1, \pm 0.3, \pm 1, \pm 3, \pm 10, \pm 30, \pm 100, \pm 300, \pm 1000, \pm 3000, \pm 10000~s
\end{equation}
\ii
Choose a set of discrete closely spaced $E_0$-values within reasonable limits, such as
\beq \label{n}
\frac {E_0} {E_a} = 1.1, 1.2, 1.3, 1.4,..., 5
\eeq
\ii
For each $n, s$ and $E_0$, a linear regression subroutine based on Eq. (\ref{e15}) furnishes $X$ and $Y$.
\ii
For each $(n,s)$ combination, evaluate
 \beq \label{tauans}
  \tau_a(n,s) = \left(\frac 1 {E_a} - \frac 1 {E_0} \right) \frac 1 Y
 \eeq
and exclude those cases for which $\tau_a(n,s)$ exceeds reasonable limits, such as those in Eq. (\ref{n'}).
 \ii
  For each $n$ and the corresponding $E_0$, evaluate the sum of squares $F(X,Y,n,s)$.
 \ii
  Finally, search for the $(n,s)$ combination that yields the smallest $F(X,Y,n,s)$.
 \ee

The vertical shift of the compliance curve is, for both cases, given by
 \beq\label{delta}
  \Del = J(t_1 - t' + s) - \hat J(t_1 - t')
 \eeq
which the $t_1$ is time of initial reporting (or any time as the vertical shift passes through whole curve).

{\bf Note:} Instead of the foregoing discrete optimization, one can formulate a continuous quadratic objective function of variables $x, y, n, s$, and impose the limits on $\tau_a$ or $E(a)$ either by the penalty method or by transformation of variables. The resulting optimization problem is highly nonlinear. It has been found that the convergence of such optimization is far slower than that of the present discrete optimization and often does not converge to the overall minimum.
The great advantage of the present discrete optimization is that it can be reduced to linear regression, which is fast, needs no initial estimates, and always gives unique result.

\section*{Verification of Filtering Algorithm and Examples}
\label{section3}

The NU database lists, for each set of experiments, the values of Young's modulus $E$ and of the mean compressive strength $\bar f_c$ of concrete at the age of 28 days. For those sets in which the experimenter did not report $E$, the database entry gives the $E$-value at age 28 days that was estimated from concrete strength $\bar f_c$ according to the ACI empirical equation (Eq. A-39 in ACI-R209 \cite{videla2008guide}). This estimate has a significant error. To convert the 28-day $E$ value to the reported age, $t_0$, at load application, the empirical equation, Eq. A-34 in ACI-R209, for the age effect on $E$ was used. This introduces an additional error.

For most of tests, the optimized values $n$ and $E_0/E_a$ are quite scattered.        
Using raw data in calibration would introduce a systematic error. Therefore, filtering of the data is necessary. Although the data in the NU database are all in terms of compliance, some or many of them may have been calculated from creep coefficient data, and it is not known which ones. So we filter all the data.      

\subsection*{Example of Algorithm Verification for Creep Coefficient}
\label{section3:sub1}

Since all the data in NU database are reported in terms of creep compliance, to demonstrate a case study for creep coefficient, we converted one set test using reported initial strain based on the following formula.
\beq \label{s111}
\vf (t,t') = EJ(t,t')-1
\eeq

\begin{figure*}
	\centering
	\includegraphics[width=0.8\textwidth]{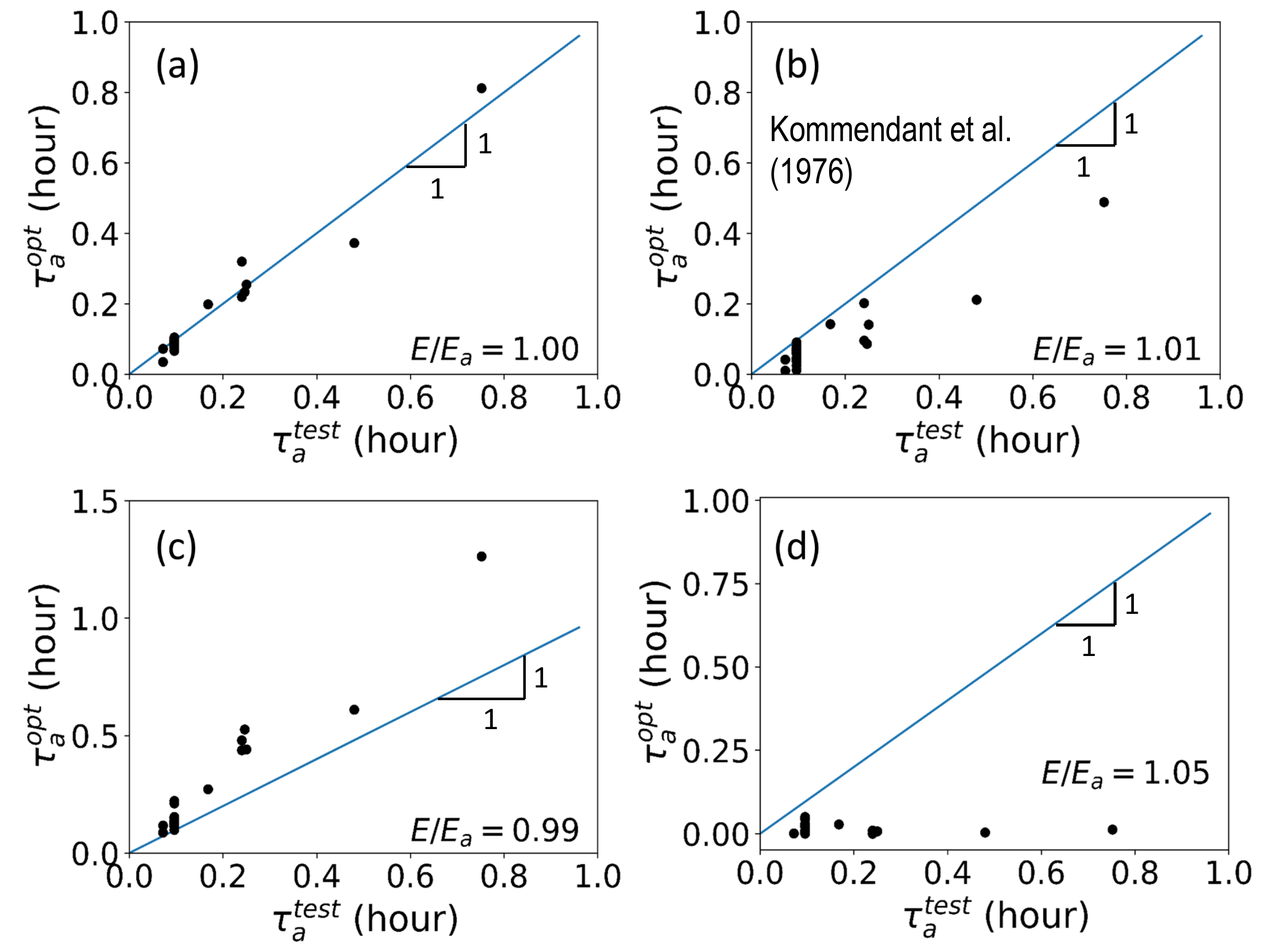}
	\caption{Sensitivity analysis of the $\tau_a$ with respect to $E$, based on Kommendant et al.'s data \cite{kommendant1976study}} 
	
	\label{fig:Sensitivity analysis}
\end{figure*}

Kommendant et al. \cite{kommendant1976study} conducted about 20 creep tests none of which is found to require any significant correction. So they are a good choice for sensitivity analysis, and particularly for verifying whether the filtering recovers the original data if they are perturbed by changing the value $E_a$. To study the algorithm for creep coefficient, the data had to be first converted to the creep coefficient. The specimens were cylinders of diameter 152 mm, tested at various stresses, ages $t'$ and temperatures. Running the present algorithm showed that no corrections were needed for these data, and even if the constraints in the algorithm were omitted. Nearly the same $\tau_a$ value was returned by the algorithm for each of the tests; see Fig. \ref{fig:Sensitivity analysis}a, which shows a plot of $\tau_a^{opt}$ versus $\tau_a^{test}$ for $E/E_a = 1$. Slope 1 of the straight line in this plot indicates a perfect agreement, and the data are close to this line, with no systematic deviation to one side or the other (these were probably the best, and best reported, creep tests ever conducted).

The agreement in Fig. \ref{fig:Sensitivity analysis}a confirms the correctness of the algorithm, and the perfect quality of data by Kommendant et al., which makes them suitable for sensitivity analysis. For this purpose, some erroneous $E$-values were introduced, the present optimization algorithm was run for each of them, and their effect on each of these tests was calculated. The results are shown in Fig. \ref{fig:Sensitivity analysis}b,c,d in terms of the optimized values of $\tau_a$ as delivered by Eq. (\ref{e8}) for each of the tests. It is now noteworthy that even changing $E/E_a$ to 1.01 or 0.99 disrupts agreement with the line of slope 1 markedly, and for 1.05 totally (ratios
$E/E_a < 1$ are, of course, physically impossible, but some testers reported data that give such ratios, and that is another reason why filtering is necessary).

\begin{figure*}
	\centering
	\includegraphics[width=0.8\textwidth]{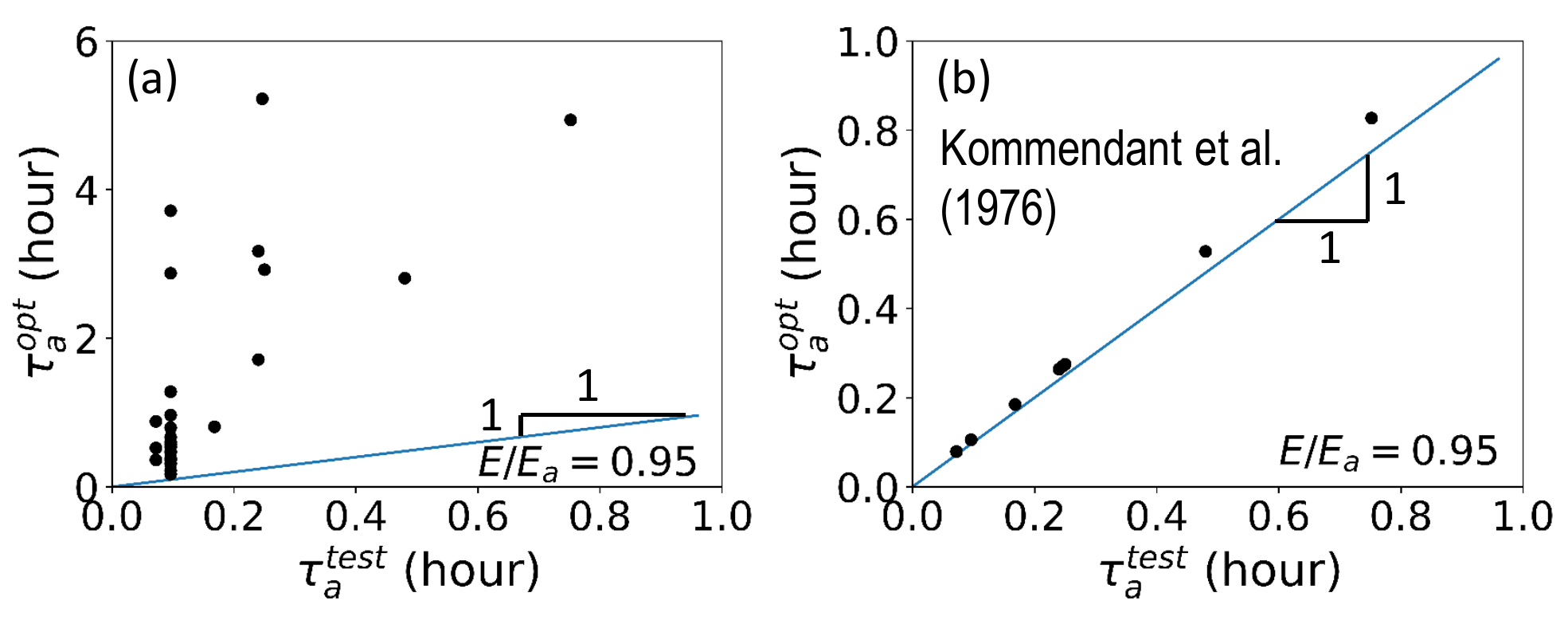}
	\caption{(a) Unconstrained and (b) constrained optimization results for deliberately altered Young's modulus $E$}
	\label{fig:Constrained Comparison}
\end{figure*}

Most importantly, the filtering algorithm is run for the same data set (Kommendant et al.), shown in Fig. \ref{fig:Constrained Comparison}a after it is disrupted by setting $E/E_a$ = 0.95. It is remarkable that, after running the filtering algorithm for the compliance function, the original values for $\tau_a$ is recovered, as shown in Fig. \ref{fig:Constrained Comparison}b. This conclusion is a powerful experimental verification which shows effectiveness of the proposed algorithm.


This example demonstrates that a measurement error in Young's modulus $E$ can generate highly distorted values for the creep coefficient and for the optimized $\tau_a$. It also shows that $\tau_a$ is very sensitive to the creep coefficient and hence is an important variable to optimize.

\subsection*{Examples of Filtering Error in Compliance Function}
\label{section3:sub2}

\begin{figure*}
	\centering
	\includegraphics[width=0.8\textwidth]{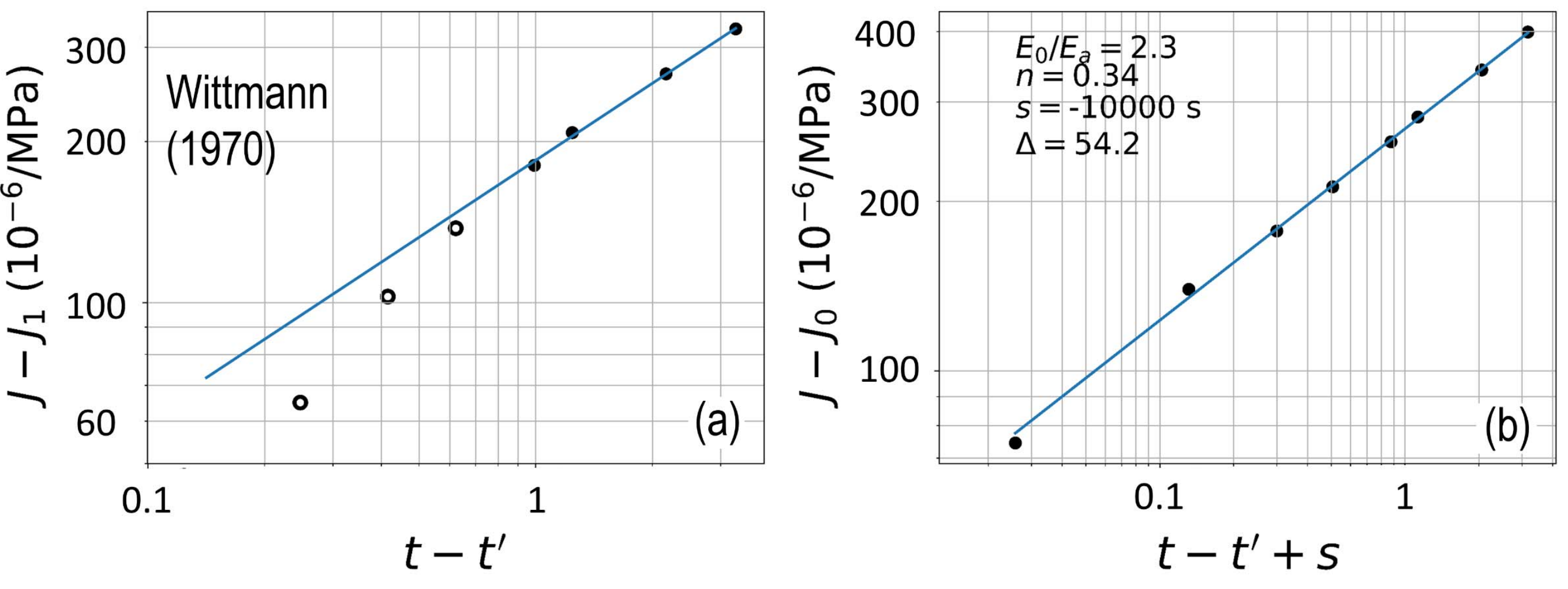}
	\caption{(a) Original (line fitted through solid points only) and (b) filtered representation of test data, reported by Wittmann \cite{wittmann1970einfluss}}
	\label{fig:Wittmann}
\end{figure*}

Fig. \ref{fig:Wittmann} shows the original and filtered representation of the creep data reported by Wittmann \cite{wittmann1970einfluss}. The specimens were cylinders of radius 18 mm and height of 60 mm, made of pure cement paste with $w/c$ = 0.40 and loaded at 80 days after casting. The raw data on creep compliance $J-J_1$ as reported are shown in Fig. \ref{fig:Wittmann}a, in the log-time scale, in which the power law is a straight line.

It is seen that only the last four data points (solid circles) conform to the power law but the earlier ones (empty circles) deviate downwards significantly. The optimization algorithm for compliance was run for these data, yielding the revised plot in Fig. \ref{fig:Wittmann}b. As seen, the filtered data are in excellent agreement with the power law and, after filtering, one finds from Eq. (\ref{delta}) the need for a significant vertical shift of the entire compliance curve, $\Del$ = 54.2 $10^{-6}$/MPa.

\begin{figure*}
	\centering
	\includegraphics[width=0.8\textwidth]{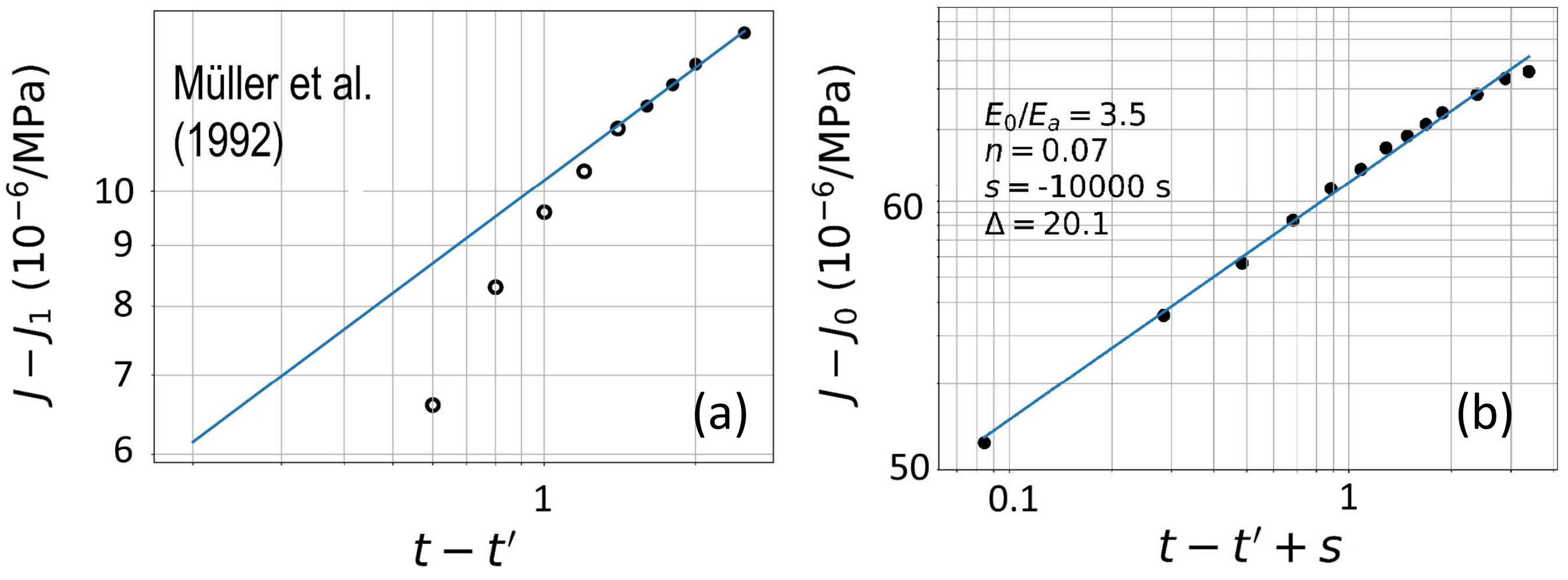}
	\caption{(a) Original (line fitted through solid points) and (b) filtered representation of the M\"uller et al. \cite{Muller1992} test data}
	\label{fig:Kuttner}
\end{figure*}

M\"uller et al.'s data \cite{Muller1992} for concrete, shown in Fig. \ref{fig:Kuttner}a for one of their 18 tests, reveal a similar behavior as Wittmann's. Again, the initial points (empty circles) deviate significantly downwards from the power law line fitted to the last five points (solid circles). This deviation for short-times is even more noticeable that for Wittmann's data the error in the ill-reported initial elastic strain, associated with the required time shift, is here comparable to the compliance value itself. The other specimens tested by M\"uller et al. exhibit similar behavior. All these specimens were loaded after 365 days, and so the power law validity should extend up to about a week. This makes these data excellent candidates to demonstrate the benefit of filtering. Indeed, the filtered data in Fig. \ref{fig:Kuttner}b show a major benefit.

The power-law for creep is best represented by a straight line in the plot of $\log(J-J_0)$ versus $\log(t-t')$. But if the time of the first data point is specified as zero, then this point cannot be plotted (as $\log 0 \to -\infty$). If that point came from a linear time scale plot, then the time assigned to this point was the shortest time discernible visually, e.g., 1 s. For the data of by Wittmann, and also of M\"uller et al., the ACI-R209 formula for estimating the elastic modulus from the strength gave a higher value than the first measured data point. In that case, the initial data point ($J_1$) was taken as the reported initial strain for the time corresponding to the first data point.

Further consider the tests by Anders \cite{anders2014stoffgesetz}. They were performed for different ages $t'$ but only two tests performed at age 28 days will be considered (for the others, $t'$ was so low that hydration intervened in the initial period). The specimens were cylinders of diameter 150 mm and height of 450 mm. The time for the first point in the database is listed as 0, which is not feasible in practice. The original report could not be retrieved.

\begin{figure*}
	\centering
	\includegraphics[width=0.8\textwidth]{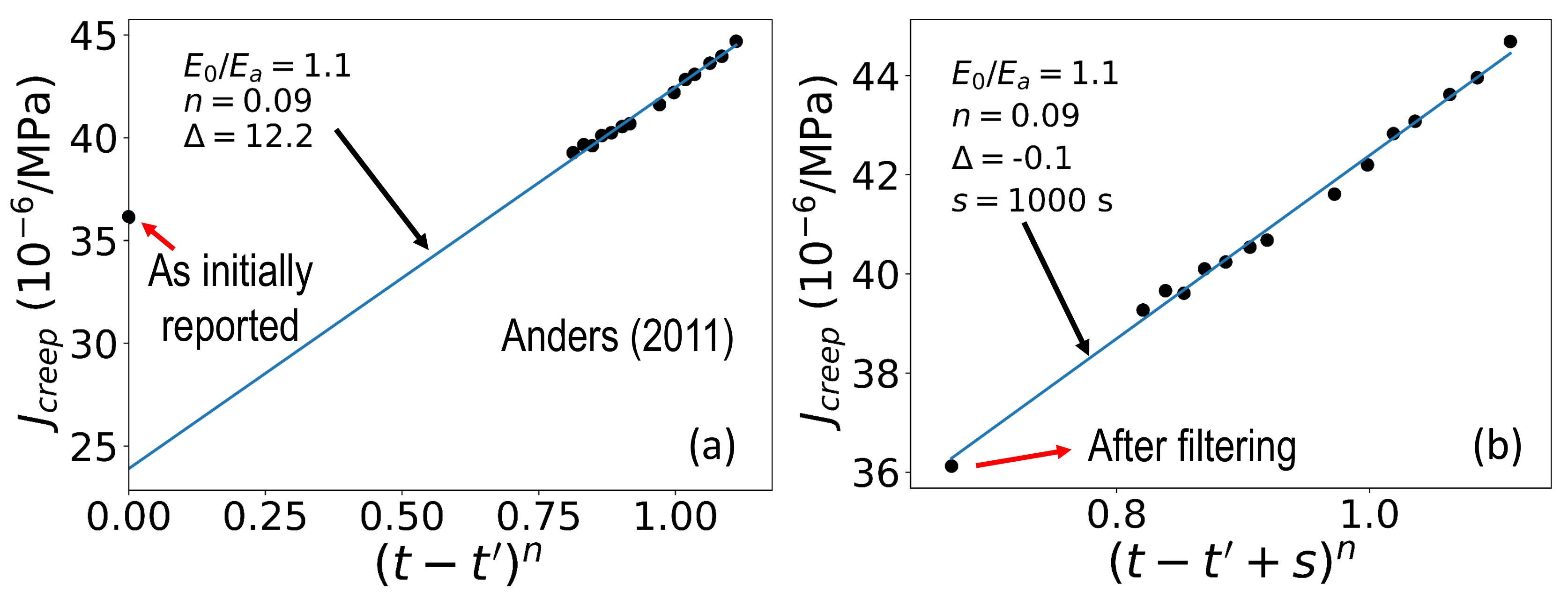}
	\caption{Optimized fit of Anders test data \cite{anders2014stoffgesetz} when time shift is (a) disallowed or (b) allowed}
	\label{fig:Anders}
\end{figure*}

So it was assumed that the time corresponding to the first point was 0.1 s, which may be the fastest loading attainable by sudden opening of a valve in a hydraulic creep frame. With this interpretation, the data are plotted in the transformed time scale in Fig. \ref{fig:Anders}.
If we no time shift is included in the optimization the time shift, a reasonably constrained optimization cannot handle these data (Fig. \ref{fig:Anders}a). But when the full optimization algorithm with time shift $s$ is applied to these data, the plot in Fig. \ref{fig:Anders}b is obtained. As seen, an excellent power-law fit is achieved, and the likely loading time for the first point is also recovered.

\subsection*{Power-Law Exponent and Its Optimum for All Short-Time Data}
\label{section3:sub3}

\begin{figure*}
	\centering
	\includegraphics[width=0.8\textwidth]{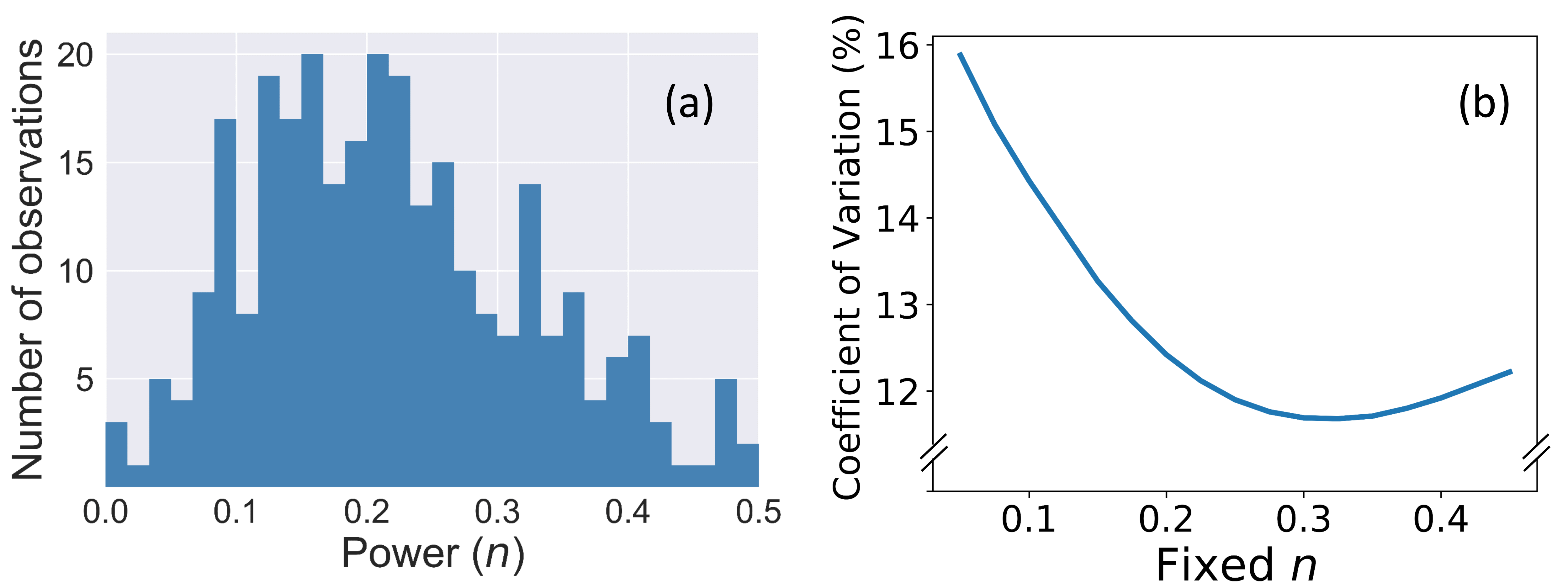}
	\caption{(a) Distribution of power law exponent $n$ in NU database, and (b) overall C.o.V. of deviations from test data as a function of $n$}
	\label{fig:fixed-n}
\end{figure*}

An unconstrained optimal fitting of the test data in the NU database showed that the $n$ value was in most tests between 0.2 and 0.3 (Fig. \ref{fig:fixed-n}a). Only in a tiny portion of the database, $n$ was less than 0.05 and more than 0.4.

Let us now determine the optimum exponent $n$ for all the tests in the database with sufficient data during the first day of sustained load, and the sensitivity of error to the value of $n$. The optimum minimizes the square of the combined coefficient of variation (C.o.V.) of regression errors, $\om_{all}$, which is calculated as
 \beq \label{s1111}
  \omega_{all} =  \sqrt {\frac{1}{N} \sum_{i=1}^{N} \omega_i^2}
 \eeq
where $\omega_i$ ($i=1,2,3...N$) is the individual coefficient of variation of the $i^{th}$ test data set in the NU database, and $N$ is total number of tests in the database. To find how $\om_{all}$ changes when $n$ deviates from the optimum, discrete values $n$ = 0.05, 0.1, 0.15, ..., 0.4 were prescribed for all the creep test series in the database, and then $\omega_{all}$ was calculated for each prescribed $n$, and plotted in Fig. \ref{fig:fixed-n}b. The optimum exponent is $n$ = 0.3, for which $\om_{all}$ = 11.1\%.

Varying $n$ does not increase the combined C.o.V of errors dramatically. Decreasing $n$ from 0.3 to 0.2 would change $\om_{all}$ by less than 2\%, and would still give almost equally good fits of the data and C.o.V increases under 2 percent (Fig. \ref{fig:fixed-n}b). So, the error sensitivity to the overall $n$ of the database is quite low. An $n$-value representing a sharp optimum can be obtained only for an individual creep test. The scatter among different concretes in the database dominates the errors and is what prevents a sharp optimum.

\section*{Five Power-Law Regimes in the Evolution of Concrete Creep}
\label{section4}

Combining the present conclusion about overall $n$ with the previous knowledge, one can discern a sequence of five regimes of power laws $\tht^n$ (where $\tht = t-t'$):
 \bea   
  \mbox{for $t-t' < $ 10 s}~~~&& n \approx 0.08 \\
  \mbox{for 10 s $\le t-t' < $ 1 day}~~~&& n \approx 0.30 \\
  \mbox{for 1 day $\le t-t' < $ 1 year}~~~&& n \approx 0.10 \\
  \mbox{for 1 year $\le t-t' < 10^3$ years}~~~&& n \to 0 \\
  \mbox{for $t-t' \ge 10^3$ years}~~~&& n = 1
 \eea
The transition between the subsequent regimes are very gradual, spread-out, and the above-mentioned times are those at which the transitions are centered.
The initial regime for $t-t'$ less than about 10 s was studied by means of creep micro-indentation tests of hardened cement paste by Vandamme et al. \cite{vandamme2013creep}. They proposed to represent their data by a logarithmic curve, which corresponds to $n \to 0$ because
 \beq \label{n to 0}
  \lim_{n \to 0} \frac{\dd}{\dd t} \left( \frac{t^n} n \right)
  = \lim_{n \to 0} t^{n-1} = t^{-1} = \frac{\dd}{\dd t}\ (\ln\, t)
 \eeq
So the creep rate for $n \to 0$ is still a power law, which again satisfies the functional equation (\ref{s1}). However, reanalysis of Vandamme et al.'s \cite{vandamme2013creep} data in \cite{wendner2015statistical} revealed that the best fit of these data is obtained by a power law with $n$ = 0.08, as stated above, although visually the difference from the logarithmic curve is quite small (later
it was proposed that the near-logarithmic creep in the ten-second nanoindentation test explains why multidecade creep is logarithmic; but this argument involved some hypotheses open to doubt).

The second regime is the present case. The third regime with $n \approx 0.10$ was identified by statistical optimization of large databases for the purpose calibration of RILEM creep prediction models B3 \cite{bavzant1995justification} and B4 \cite{bazant2015rilem} and was incorporated into a general expression of $\dd J(t,t')/ \dd t$, which also covers the transitions between the regimes with $n$ = 0.10 and 0.

The fourth regimes, $n \to$  0, is the long-time creep evolution, which is logarithmic, as reflected in Eq. (\ref{n to 0}). For all practical purposes, it is the terminal regime of long-time concrete creep, as proposed in 1975 in \cite{bazant1975creep} and first incorporated into engineering society recommendations in 1995 (RILEM model B3). As a debatable point, it may be mentioned that some researchers \cite{vandamme2013creep} suggested the logarithmic multidecade creep to be a necessary consequence of a supposedly logarithmic creep seen in the aforementioned 10-second nanoindentation experiments (which itself is more likely a power law, as mentioned above).

The fifth and last regime is purely speculative and, for the service stress range of concrete, experimentally unverifiable. It is irrelevant for structural engineering practice. But it is interesting theoretically, e.g., because of analogy with the creep shale \cite{chau2017enigma} and other rocks over geologic times.
In materials science, it is generally accepted that the primary creep (which is all that applies to concrete practice) eventually transits to a secondary creep, which is constant-rate viscous flow and depends on the sustained stress nonlinearly. The necessity of this transition may be explained by the fact that (except in perfect crystals) every deformation increment must activate new creep sites (i.e., sites of nanoscale slip at overstressed locations in the microstructure; see \cite[Fig. 6]{chau2017enigma}). The gradual transition between the primary and secondary creep is centered at the so-called Maxwell time \cite{chau2017enigma}, which has been roughly estimated from a certain geological evidence to be between $10^4$ and $10^6$ years for rocks such as shale, and is probably shorter for concrete. The secondary creep of concrete should be observable at normal times only under very high shear stress combined with hydrostatic pressure high enough to prevent all microcracking.

The physical reasons for the changes in power-law exponent through subsequent time periods deserve further study.

\section*{Conclusions}
\label{con}

   \begin{enumerate}
	\item
The correct time scale for graphical presentation of the test data on creep (as well as shrinkage) is the logarithmic scale. Reporting in the linear scale of time (or test duration) is grossly misleading.
    \item
The best way of reporting concrete creep data is a numerical table (nowadays on a computer), in which the measurement times should be refined in an approximately geometric progression as the test duration decreases; e.g., $t-t'$ = 0.1, 0.316, 1, 3.16, 10, ..., 1,000, 3,160, 10,000 s.
	\item
For a short initial period during which the changes in the effects of hydration and selfdesiccation (and possibly also of drying) are negligible, a scaling analysis shows that the creep strain must initially evolve as a power law.
    \item
Most reported test data are imperfect and need corrections. The time entering the compliance function needs to be corrected by a certain time shift due to the fact that the process of load application takes a finite time, ranging from 0.1 s to even several hours, which is a circumstance usually unreported. The time shift is optimized such that the reported initial strains would follow a power function of time. The time shift produces a vertical shift of the entire compliance function curve.
    \item
The creep coefficient values as well as the compliance function values are rather sensitive to Young's modulus $E$ which, in turn, varies greatly with the time, $\tau_a$ (needed to apply the load and the load application history). Mixing data with very different $\tau_a$-values introduces large errors into the creep database. Their correction leads to a vertical shift of the compliance function curve.
	\item
Based on the foregoing conclusion, the way to filter out the errors from the database is to use statistical optimization so as to shift the time and strain scales with the objective of making the initial creep data fit a power function of time as closely as possible.
    \item
The sum-of-least-squares optimization problem is reducible to a linear regression that is subjected to several nonlinear constraints, which are best tackled in a discrete form.
	\item
Enforcing a power law form of the initial creep curve generally leads to better fits of test data.
    \item
The optimum exponent $n$ of the power law for initial creep varies from one data set to another, but is mostly between 0.10 and 0.35. The optimum for the whole database, minimizing the coefficient of variation of errors, is $n \approx 0.3$.
    \item
After filtering out the errors, the corrected database will allow better calibration of the creep prediction model such as B3 or B4.
   \end{enumerate}
 	

\textbf{Funding}: This study was partially funded by DoE through Grant 20778 from the Infrastructure Technology Institute of Northwestern University, and from the NSF under grant CMMI-1129449.

\textbf{Compliance with Ethical Standards: } Compliance is certified.

\textbf{Conflict of Interest}: The authors declare that they have no conflict of interest.

\bibliographystyle{abbrv}
\bibliography{ReferencesShr}






\end{document}